\documentclass[journal]{IEEEtran}
\usepackage[dvips]{graphicx}
\usepackage{indentfirst}
\usepackage{amsmath}
\usepackage{amssymb}
\usepackage{threeparttable}
\usepackage{multirow}
\usepackage{subfigure}
\usepackage{xcolor}
\usepackage{cite}
\usepackage{enumerate}
\usepackage{algorithm}
\usepackage{algorithmic}
\usepackage{setspace}
\usepackage{multicol}
\usepackage{stfloats}
\usepackage{array}

\ifCLASSOPTIONcompsoc
\else
\fi

\ifCLASSINFOpdf
\else
\fi

\allowdisplaybreaks

\begin{document}

\title{Integrated Communication, Navigation, and \\ Remote Sensing in LEO Networks  \\  with Vehicular Applications}

\author{

Min Sheng,~\IEEEmembership{Senior Member,~IEEE},
Chongtao Guo,~\IEEEmembership{Member,~IEEE}, and
Lei Huang,~\IEEEmembership{Senior Member,~IEEE}

\thanks{
M. Sheng is with the State Key Laboratory of Integrated Service Networks, Xidian University, Xi'an 710071, China (E-mail: msheng@mail.xidian.edu.cn).
C. Guo and L. Huang are with the State Key Laboratory of Radio Frequency Heterogeneous Integration, College of Electronics and Information Engineering,  Shenzhen University, Shenzhen 518060, China (E-mail: ctguo@szu.edu.cn; lhuang@szu.edu.cn).
C. Guo is the corresponding author.
}
}

\maketitle

\vspace{-6em}
\begin{abstract}
Traditionally, communication, navigation, and remote sensing (CNR) satellites are separately performed, leading to resource waste, information isolation, and independent optimization for each functionality.
Taking future automated driving as an example, it faces great challenges in providing high-reliable and low-latency lane-level positioning, decimeter-level transportation observation, and huge traffic sensing information downloading.
To this end, this article proposes an integrated CNR (ICNR) framework based on low Earth orbit (LEO) satellite mega-constellations.
After introducing the main working principles of the CNR functionalities to serve as the technological basis, we characterize the potentials of the integration gain in vehicular use cases.
Then, we investigate the ICNR framework in different integration levels, which sheds strong light on qualitative performance improvement by sophisticatedly sharing orbit constellation, wireless resource, and data information towards meeting the requirements of vehicular applications.
We also instantiate a fundamental numerical case study to demonstrate the integration gain and highlight possible future research directions in managing the ICNR networks.
\end{abstract}


\section{ Introduction }

\IEEEPARstart{I}{n} the past thirty years, the ever growing human demand on seamless information coverage, terminal positioning, and earth observation makes global artificial satellite industry evolve and expand  significantly, leading to three main independent bodies: communication, navigation, and remote sensing (CNR) satellite systems.
Due to different focuses, the existing satellites are operating in various orbits, including geo-stationary Earth orbit (GEO) with an altitude of 35,786 km, medium Earth orbit (MEO) with an altitude of from 2,000 km to 20,000 km, and low Earth orbit (LEO) with an altitude of less than 2,000 km.
Specifically, the UK OneWeb communication systems, the USA global positioning system (GPS), and China Gaofen 4 remote sensing satellites are working in  LEO, MEO, GEO orbits, respectively [1].

While deciding the altitude of a single satellite, there is a fundamental tradeoff between coverage area and signal propagation latency.
However, by developing a huge constellation of low-altitude satellites, the coverage can be enhanced while reducing latency, which brings about a rising tide of world-wide LEO satellite network construction in the past ten years.
Meanwhile, technological advancements have made it possible to build small cost-effective LEO satellites, develop reusable rockets, and adjust moving status in orbit.
In this context, the economically feasible LEO satellite manufacture, launching, and in-orbit management techniques have enabled many countries, organizations, and commercial companies to establish their own  satellite constellations.
Besides the advantage of low transmission latency and small signal attenuation, the huge constellation of LEO satellites is also good at enhancing capacity and reliability of communication, accuracy and rapidity  of positioning, and spatial and temporal resolution of remote sensing [2].
Due to such a great potential, the number of LEO satellites may persistently grow towards the capacity of the orbits, predicted as several millions, leading to mega-constellations of LEO satellites in the future [3].

It had long been the situation that different kinds of satellites usually serve the customers with distinct purposes.
Particularly, communication satellites mainly provide special users connectivity in regions out of terrestrial network coverage, such as sea, forest, and desert  regions.
Navigation satellites offer global positioning, navigation, and timing (PNT) services for a large body of location-dependent applications, including outdoor expedition, drone management, and industrial automation.
Remote sensing satellites are able to provide a top view of the earth, which can be applied to agriculture monitoring, natural disaster discovering, and traffic network optimization.
However, heterogeneous services can be required by vehicular terminals, especially in the era of autonomous driving.
In particular, an automated connected vehicle  in intelligent transportation systems may require timely remote-sensing traffic  information to optimize its route, necessitate high-precision positioning to perform lane-level navigation, and need  high-capacity communication to transfer diversified messages including remote sensing data [4].
To efficiently provide comprehensive CNR services, the LEO mega-constellation of satellites is expected to be a promising solution.

\begin{figure*}[!t]
\centering
\includegraphics[width=0.99\textwidth]{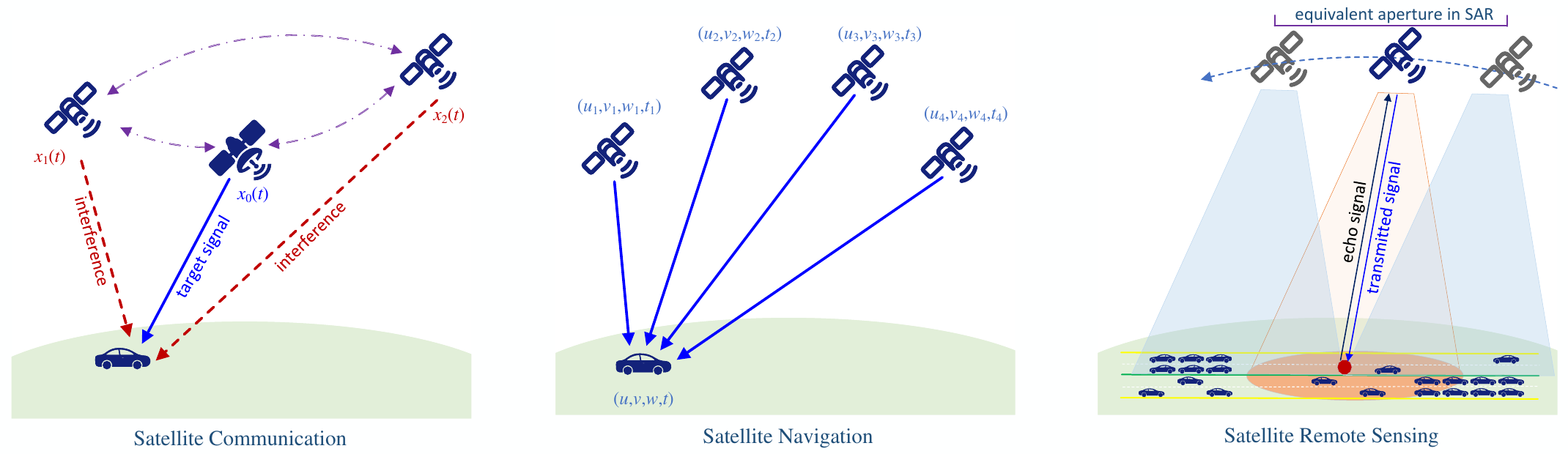}
\caption{Fundamental working principles  of satellite communication, navigation, and remote sensing. \label{Fig:1}}
\end{figure*}

Traditional separately deployed CNR satellite systems face considerable challenges in meeting captious vehicular requirements  due to loss of sharing gain in terms of orbit constellation,  wireless resource, and data information, which further prohibits joint CNR optimization towards satisfying the autonomous driving application.
In this regard, the framework of integrated communication and navigation has been proposed.
In the early stage of such integration dating back to 1980s, high-altitude satellites are mainly designed for vehicular navigation, implying that navigation is taken as the primary function and communication is introduced as the secondary capability.
With the proliferation of LEO satellite based broadband Internet, the integration moves the foundation to the communication side, where navigation data is incorporated into the communication packets [5].
Another topic is the integration of communication and remote sensing, which has hardly been considered until the recent flourishing low-altitude space networks, mainly because the delay requirement of remote sensing data downloading for traditional users is not as strict as present applications, such as intelligent transportation.
Particularly,  communication among remote sensing satellites offers the opportunity of onboard sensing information fusion, leading to higher resolution and quicker response, which is pretty beneficial to modern vehicular users [6].
Other than the combination of dual functionalities, a few works have begun to consider the combination of the three functionalities.
Frequency diverse array and orthogonal frequency division multiplexing have been utilized in [7] to efficiently exploit waveform diversity for integrated communication, sensing, and navigation.
In [8], the constellation design has been investigated and a simplified analysis has been provided to jointly consider the communication and remote sensing requirements.
The attractive prospect of the integration with applications in ecological environment monitoring has been demonstrated in [9].
Last but not least, the national aeronautics and space administration (NASA) has also raised its interests on multi-purpose space cubes with remote sensing, communication, and navigation integration [10].

Despite  sporadically mentioned in the  literature,  the integration of the CNR in a mega-constellation of LEO satellite networks has not been completely investigated with its fundamentals, classification, potentials, and challenges being specified.
In this article, we first summarize the most fundamental principles of the CNR techniques individually, followed by demonstration of the proposed integrated CNR (ICNR) framework with vehicular applications as a representative use case.
Then, different levels of integration in the proposed ICNR network are discussed and a fundamental numerical case study is conducted to demonstrate the integration gain.
Finally, we touch on possible future research directions that need to be elaborated in practical  network deployment and management.

\section{ ICNR Framework with Vehicular Applications}

In this section, we  first present the fundamental principles of the satellite CNR systems, as described in Fig. 1.
Then, an ICNR framework is proposed with typical use cases illustrated in vehicular applications.

\subsection{Satellite Communication}

Traditional communication satellites usually act as relays to transit information, i.e., forwarding information from a source node to a destination node on the earth, where the destination node  can be as far as  thousands of kilometers away from the source node.
Taking downlink transmission as an example, we consider a terminal  trying to capture information signal, $x_0(t)$, from a host satellite in Fig. 1.
In the zenith of the terminal, there usually exist some additional satellites emitting specified signals, such as  $x_1(t)$ and $x_2(t)$, to their associated users, causing interference to the terminal.
In this regard,  the terminal  generally receives  delayed and attenuated signals from the host satellite and the interfering satellites as well.
The main task of satellite communication is to recover the target signal  from the superimposed signal by removing the interference and noise.
Generally, capacity, reliability, and latency are the three dominant metrics that are usually cared about by vehicular users, e.g., to download the remote sensing data.
A peculiarity of vehicular applications here is the mobility of terminals, which makes channel features to be different from static users and deserves special attention in practical systems.

\subsection{Satellite Navigation}

Positioning and timing are the two main tasks of navigation systems, which are the most fundamental requirements of vehicular terminals.
Navigation satellites can usually be aware of time and their trajectories, with the aid of  clocks equipped and periodical ephemeris acquisition from ground observation stations.
Consider that a target user aims to determine the time $t$ and its position denoted by three-dimensional coordinates $(u,v,w)$ in a Cartesian coordinate system.
To this end, upon the user receiving a signal from a reference satellite, it is able to get the distance from itself to the reference satellite by multiplying the speed of light and the time it takes for the signal to travel.
Such a distance can also be expressed by calculating the Euclidean distance between the positions of the target user and the reference satellite using simple geometry.
Then, with one reference satellite in hand, we have one distance equation with four unknown variables.
To determine these four unknowns, at least four reference satellites are required to establish four independent distance equations, given that the position and clock time of the $i$th reference satellite, jointly denoted by $(u_i,v_i,w_i,t_i)$, are konwn.
Due to ranging errors caused by inaccurate reference satellite position, multi-path propagation, light refraction, etc., the position and time are estimated rather than computed exactly.
Generally, receiving signals from additional satellites does positioning and timing a favor with more precision  by finding a least-square solution of the distance equations with more than four available navigation satellites.
Besides the number of visible satellites, the relative geometric relationship among user equipment and visible satellites also plays  a significant role in determining the accuracy of positioning and timing,
usually measured by the geometric dilution of precision (GDOP) indicating the ratio of positioning and timing error to ranging error [11].
Vehicular terminals can also leverage inertial measurement units (IMU) and cooperative technique to improve the positioning accuracy in addition to exploiting available satellite navigation signals.

\begin{figure*}[!t]
\centering
\includegraphics[width=0.98\textwidth]{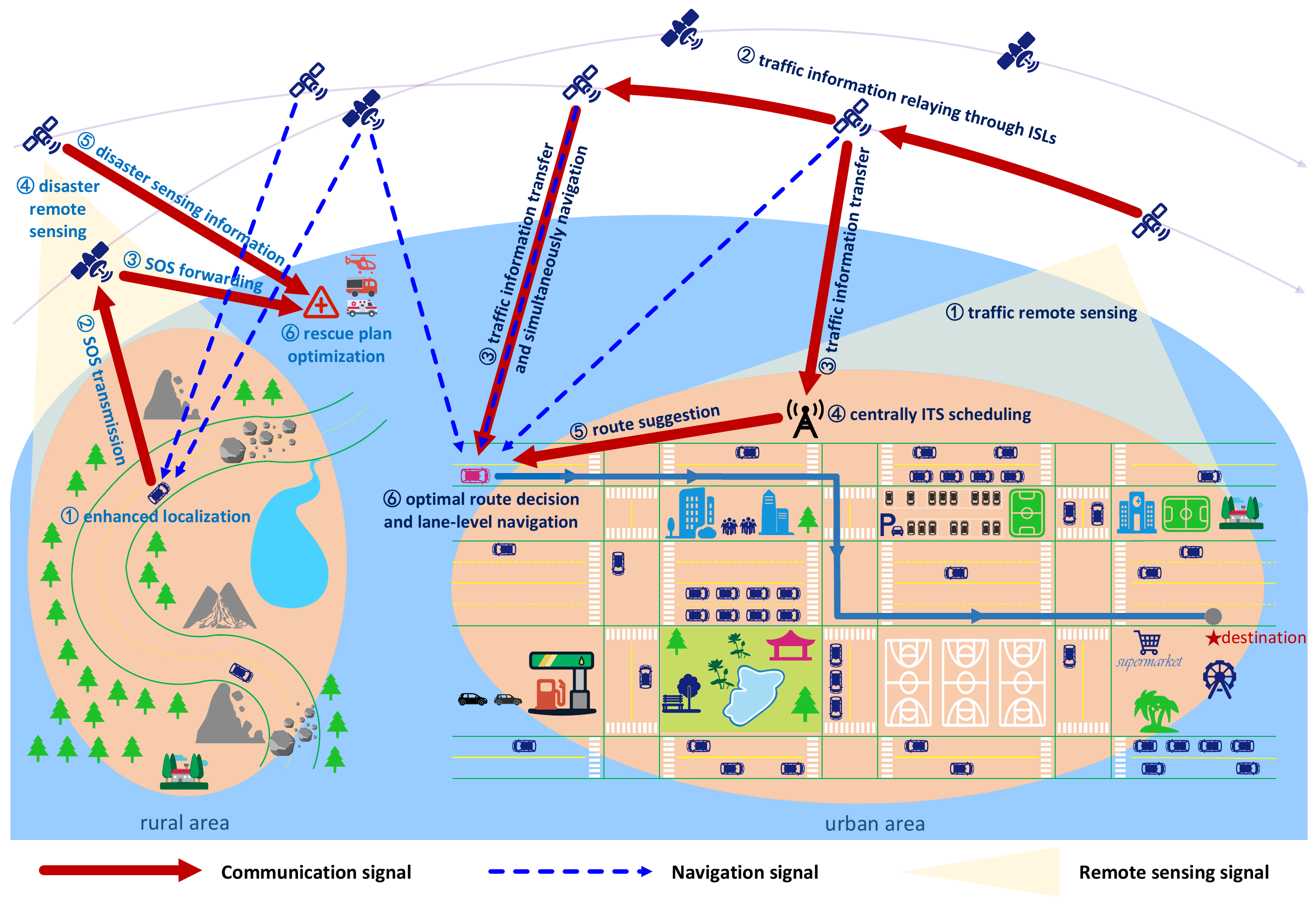}
\caption{ Illustration of the proposed ICNR framework  with vehicular applications considered in representative rural and urban use cases. \label{Fig:2}}
\end{figure*}

\subsection{Satellite Remote Sensing}

Remote sensing satellites are  designed to view the earth from a position in orbit and usually produce images of the earth.
These images are crucial to a diverse array of applications, including environmental monitoring, weather prediction, agriculture management,  disaster detection, and intelligent transportation.
There are generally two kinds of satellite remote sensing technologies, namely passive and active ones.
The passive remote sensing satellites detect energy emitted or reflected from objects on the earth, usually operating in visible, infrared, thermal infrared, and microwave portions of the electromagnetic spectrum.
Since the passive sensing technology is vulnerable to lighting conditions and dependent on ambient energy, the active remote sensing, especially synthetic aperture radar (SAR) based remote sensing,  plays an important role in mission-critical applications.
Similar to general radar remote sensing mechanism, a SAR satellite performs electromagnetic radiations to illuminate the target on the earth and subsequently records the reflected radiation signal to get earth observations.
Due to limitations on antenna size and waveform frequency, it is generally difficult to capture high-resolution images of the surface of Earth.
However, by exploiting the motion of spacecrafts, SAR satellites can achieve equivalent long-aperture antennas by accumulating signal reception over time, leading to a  resolution to be a half of the antenna length in the azimuth direction.
Although a SAR satellite prefers a small length of the antenna,  practical antenna size  should be large enough to restrict the illuminated ground area to avoid ambiguous echo signal reception [12].
With the pulse compression technique, the cross-track resolution of a SAR satellite is proportional to the speed of light divided by the product of the spectrum bandwidth and the sine of the view angle, where the view angle is the satellite centered angle between the ground target and the nadir.
In vehicular use cases, the spatial resolution should be high enough to distinguish different vehicles.
Besides the spatial precision, the temporal precision also deserves special attention, which is traditionally captured by the revisit time.
It is worth noting that the temporal resolution should be defined with respect to the whole system in a mega-constellation of LEO satellites, i.e., the time it takes for a target to be observed twice by the constellation rather than by a satellite.
The temporal resolution needs to be comparable to the variation period of the road traffic in vehicular applications, which can be measured by age of information (AoI) of the remote sensing data received by vehicles [13].

\begin{figure*}[!t]
\centering
\includegraphics[width=0.95\textwidth]{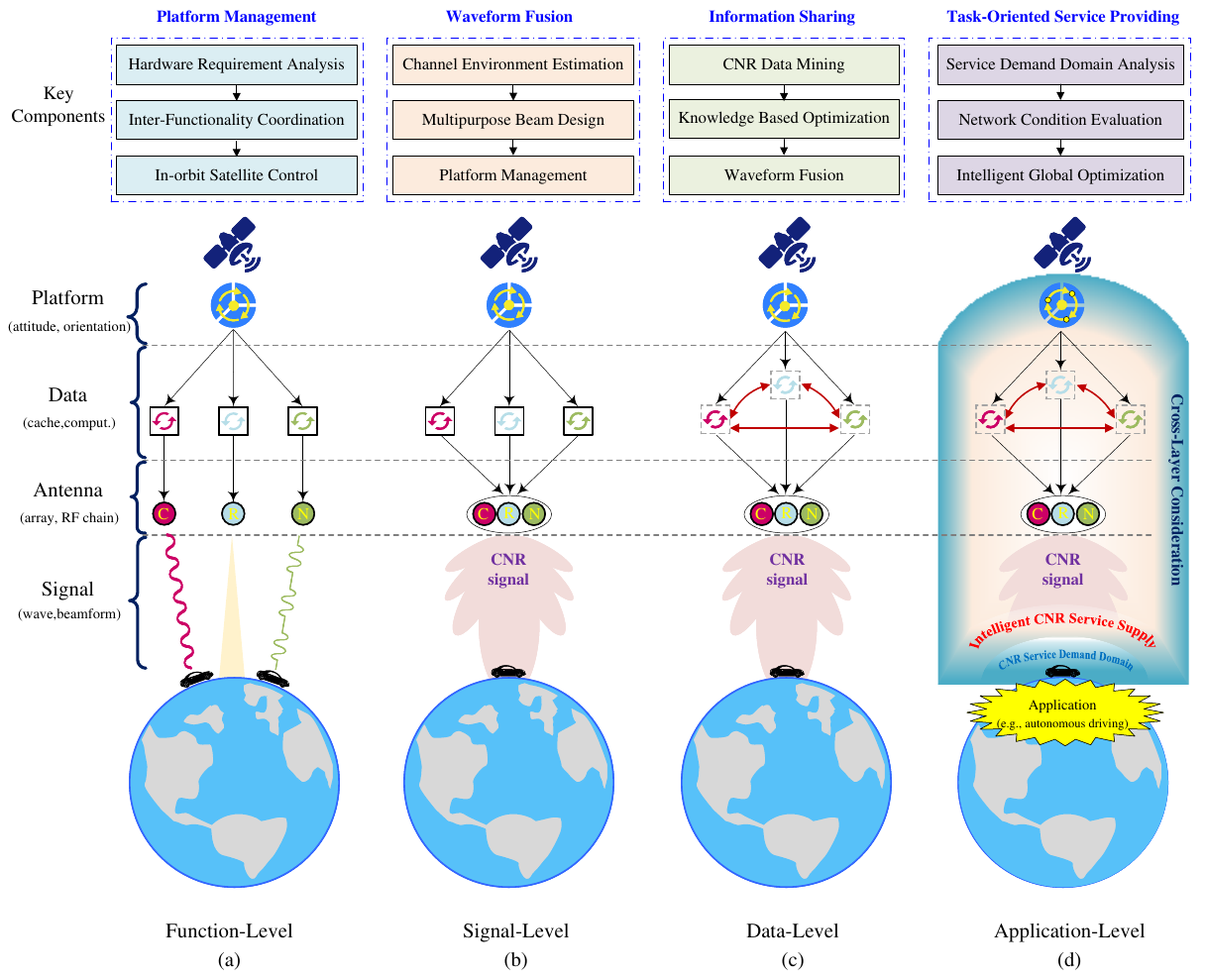}
\caption{Key components and working mechanisms of four different integration levels of the proposed ICNR framework. \label{Fig:3}}
\end{figure*}

\subsection{Proposed ICNR Framework}

Since all of the CNR satellites involve electromagnetic signal transmitting and receiving, we propose a LEO satellite based ICNR  framework by integrating the three distinct functionalities in a single satellite and constructing an ICNR network with inter-satellite links (ISLs).
Such an ICNR system is operationally feasible thanks to ever developing satellite manufacturing and launching technologies.
Actually, the Iridium-next system is planning to embrace communication payload, navigation enhancement payload, and earth observation sensors within a single satellite [14].
More importantly, orbit, spectrum, time, and energy resource can be more efficiently exploited compared with the traditional physically separate and independent systems.

Particularly, operating in a mega-constellation LEO satellite networks, the ICNR framework sheds new light on rural and urban vehicular applications, as illustrated in Fig. 2.
In case of being trapped on roads due to natural diaster in distant regions without terrestrial infrastructure coverage, the vehicle can first get its enhanced localization by exploiting more visible LEO satellites and then send a  save-our-ship (SOS) signal to an available satellite, which further forwards the SOS signal to a ground rescue station.
On the other hand, an appropriate satellite can be triggered to perform real-time remote sensing for the related region and then send  the disaster sensing information to the ground rescue station.
Based on the geographical disaster condition and the location of the stranded vehicle, the best rescue plan can be made, e.g., determining the best transportation means and the rescue trajectories.
The ICNR system is also expected to perform remote driving operation monitoring and cooperative perception in rural areas with the aid of high-bandwidth satellite communications.

The ICNR networks are also quite beneficial to future intelligent transportation systems (ITS) even within the coverage of terrestrial networks, as presented in the urban area in Fig. 2.
Because of stochastic traffic distribution, the shortest driving path may not be the best choice for a vehicle.
In this context, a satellite may first implement remote traffic sensing for the area covering both the current location and the destination point.
Then, such sensed traffic information is forwarded through ISLs to other satellites, which are closer to the vehicle or the ITS scheduling center.
After that, a proper relaying satellite can transfer the traffic distribution information to the vehicle to let it make a better route plan.
Alternatively, a relaying satellite may also send the  traffic distribution information to the ITS scheduling center, which produces a route suggestion after considering the whole traffic system and notifies the vehicle.
Although terrestrial ITSs can deploy a massive number of sensors on roads, it is quite difficult for vehicles to periodically get synchronous traffic information from these sensors due to limited communication resource. Even such information can be timely collected, it is still challenging to depict a whole picture of the road traffic because heterogeneous sensors produce multi-mode data and different equipment have different angles of view. By contrast, the proposed ICNR system is able to observe the interested transportation area as a whole and outputs a high-resolution remote sensing image, which can be downloaded to vehicles by high-capacity communication. Furthermore, aircrafts in the low sky, e.g., unmanned aerial vehicles (UAVs), can also be deployed as complementary platforms to perform remote traffic sensing for the regions that are not directly visible to satellites.
Other than the traffic distribution information acquisition, it becomes feasible for the ICNR system to achieve more enhanced centimeter-level positioning by exploiting more visible LEO satellites and simultaneous multi-purpose signal transmission.
Finally, based on the timely traffic congestion information, the accurate positioning, and the suggestion from the ITS center, vehicles are able to make the optimal route decision and lane-level navigation.
In summary, real-time vehicle-distinguishable remote sensing, lane-level positioning, and huge remote sensing information downloading are commonly required by intelligent connected vehicles, which may be suitably addressed by the proposed ICNR networks with a mega-constellation of LEO satellites.
Particularly, the superiority becomes more pronounced in the case of  ground network congestion and terrestrial emergency even in urban areas.

\section{Integration Levels of ICNR}

In this section, we classify the proposed ICNR framework into four stages, i.e., function-level, signal-level, data-level, and application-level integration, in an increasing order of the coupling degree, where  key system components and  working mechanism of different integration levels are summarized in Fig. 3.

\vspace{0.5em}
\begin{itemize}
\item L1: Function-Level Integration
\end{itemize}
\vspace{0.5em}

Function integration stands for the lowest integration level of the ICNR networks, where the CNR functionalities are loaded in a  single LEO satellite but operate almost independently.
Each functionality uses its exclusive hardware  to accomplish its own task, except sharing the same orientation and attitude with respect to each satellite.
This means that the ICNR networks can be sometimes treated as three mutually independent systems for the CNR, with a common constellation shared.
The key system module in such a level of integration is the platform management, including components of hardware requirement analysis, inter-functionality coordination, and finally in-orbit satellite control.
As the equivalent number of satellites for each functionality increases by such integration, the CNR performance can be overall improved in terms of communication coverage, positioning accuracy, and remote sensing response time.
A typical potential of such integration is to leverage the broadband communication capability to perform resolution enhanced multi-satellite sensing fusion and low-latency  transportation sensing information downloading  for automated driving applications.

\vspace{0.5em}
\begin{itemize}
\item L2: Signal-Level Integration
\end{itemize}
\vspace{0.5em}

In this level, the signals for the CNR are expected to be combined, i.e., the signal emitted by a satellite carries communication information, navigation reference, and remote sensing probes simultaneously, which leads to a more efficient use of limited spectrum resource than the function-level integration.
As a core system module, waveform fusion can be achieved by channel environment estimation and multipurpose beam design, based on the platform management as demonstrated in the function-level integration.
We can also have an easier variant that  allocates different functionalities orthogonal wireless resource by division of time, frequency, or space.
Generally, there is a fundamental tradeoff among different performance of the CNR in designing multi-purpose signals.
In practical systems, how  functionalities are combined in the signal should depend on the real-time geographical distribution of the CNR requirements.
Particularly, for a time instance at which only communication and navigation services are required by the ground users in a region, there is no need to embrace remote sensing capability in the current signal design.
The vehicular terminals can benefit greatly from equivalent enlarged resource pool with signal-level integration, which holds attractive potentials to provide high-resolution traffic sensing information and lane-level navigation.

\vspace{0.5em}
\begin{itemize}
\item L3: Data-level Integration
\end{itemize}
\vspace{0.5em}

Different from the above two levels, data-level integration allows the satellites to share information collected by different functionalities, where the information coming originally from one functionality may be very helpful in optimizing the network performance from the perspective of another functionality.
The most prominent system module here is the information sharing, which is composed of three leading components, i.e., CNR data mining, knowledge based optimization, and waveform fusion.
For instance, the remote sensing results generally show the geographical vehicular traffic distribution, which can be leveraged to conduct more efficient traffic-aware communication and navigation resource scheduling.
Using navigation information, such as location and speed of vehicles, the communication channel may be more efficiently predicted, which allows more adaptive communication transceiver design.
By analyzing the communication channel state information (CSI) of a large number of vehicular users, we can estimate the overall channel conditions  for a given region, which can be exploited to perform better remote sensing beamforming design and CSI fingerprinting based positioning.
Finally, high mobility of ground vehicles causes rapid change of  remote transportation sensing images, navigation GDOP values, and communication load distribution, which gives rise to a high demand on fast inter-functionality data mining.

\vspace{0.5em}
\begin{itemize}
\item L4: Application-Level Integration
\end{itemize}
\vspace{0.5em}

All  information delivery, terminal localization, and earth  observation functionalities are actually supporting technologies rather than the final objectives of the ICNR system.
To this end, the application-level integration aims to support  vehicular tasks, rather than meet predefined CNR requirements as in the other three levels of integration.
In particular, the CNR service requirements raised by a specific application can be expressed as a multi-dimensional domain, where different performance aspects can be somewhat interchangeable.
Taking autonomous driving as an example, the vehicle may require periodic high-resolution traffic sensing data to perform adaptive route planning, or alternatively more frequent acquisition of traffic information with less accuracy, i.e., the requirements on communication and remote sensing can be interchangeable.
Based on the real-time system condition, such as satellite geometry, channel environment, and user distribution, the application-level integration can choose an appropriate service target within the CNR demand domain,
which will be provided by intelligent global optimization with cross-layer joint consideration of information sharing, waveform fusion, and platform management.
Therefore, towards supporting various vehicular applications, the system requires a task-oriented service providing module, consisting of service demand domain analysis, network condition evaluation, and  intelligent global optimization components.
The intelligent service supply adaptive to demand domain and network condition  holds great potentials to satisfy a massive number of vehicular users with diversified applications.

\begin{table}
\centering
\begin{footnotesize}
\centering
\caption{Network Setup of the Case Study}\label{Table.I}
\renewcommand\arraystretch{1.3}
\begin{tabular}{|m{0.51 \linewidth}|m{0.39\linewidth}|}
\hline
{\textbf{Requirement/Parameter/Scenario}}    & \textbf{Value or Description} \\\hline
Numbers of traditional CNR satellites        & $4500$, $250$, $250$  \\\hline
Bandwidth of traditional CNR satellites      & $250$, $25$, $25$; MHz    \\\hline
Communication transmit power                 & $20$ W  \\\hline
Navigation transmit power                    &  $20$ W \\\hline
Remote sensing transmit power                & $80$ W  \\\hline
Driving area                                 & Remote rural region without terrestrial coverage  \\\hline
Vehicle density                              & Very sparse without inter-vehicle connection  \\\hline
Velocity of vehicles                         & Less than 108 km/h  \\\hline
Perception and maneuver reporting rate       & Once every 3 meters  \\\hline
Perception data size                         & 100 detected objects with 80 bytes for each [15]   \\\hline
Maneuver data size                           & 500 bytes [15]  \\\hline
Navigation ranging error of unit distance    & $2.28\times 10^{-8}$ m \\\hline
Communication elevation angle                & $60^{\circ}$ \\\hline
Navigation elevation angle                   & $20^{\circ}$ \\\hline
Remote sensing view angle                    & $30^{\circ}$ \\\hline
Remote sensing swath width                   & $20$ km \\\hline
Communication minimum required rate          & $0.7$ Mbps \\\hline
Remote sensing data size                     & 360 Gbits \\\hline
Remote sensing data delivery rate            & 1 bps per Hz \\\hline
Large-scale fading                           & $110 + 37.6 \log_{10}({\rm distance})$, ${\rm distance}$ in km  \\\hline
Small-scale fading                           & Rayleigh fading  \\\hline
\end{tabular}
\end{footnotesize}
\end{table}

\section{A Fundamental Case Study}

This section evaluates performance gain achieved by the lowest two levels of integration for autonomous driving in rural area without terrestrial network coverage.
Each type of  satellites is uniformly distributed on the sphere with an altitude of 500 km.
In the case of function-level integration, the number of the ICNR satellites is set as the sum of the numbers of the traditional CNR satellites.
One step further, we assume ideal waveform design such that all the CNR functionalities can share the overall bandwidth of the traditional CNR satellites  without performance tradeoffs in the signal-level integration.
There are 1,400 vehicles sparsely distributed on circular cap area of $7.36 \times 10^5 \ {\rm km}^2$, moving with a speed of less than 108 km/h.
Each vehicle reports its coarse sensing information and driving maneuver intention to LEO satellites once every 3 meters, such that the ITS center can monitor the driving performance and also provide cooperative perception service to others.
With the perception parameters adopted in [15], such periodical message delivery roughly requires a safeguarded data rate threshold 0.7 Mbps, and communication outage occurs if achievable data rate is lower than the  threshold.
Besides such safety-related message transfer, vehicles also requires best-effort downloading service to provide passengers with online infotainment, which cares more about the metric of ergodic capacity.
Meanwhile, every user needs positioning, timing, and remote sensing information, where the LEO enhanced positioning is performed on the basis of the GPS.
In the integration cases, we consider each swath is illuminated by two satellites simultaneously to get observation resolution improvement, say, by a half, in the range direction by data fusion.
The AoI of the remote sensing data received by the ground users is derived by adding satellite revisit time and the data downloading time.
Some key simulation parameters are summarized in Table I.

\begin{figure}[!t]
\centering
\includegraphics[width=0.46\textwidth]{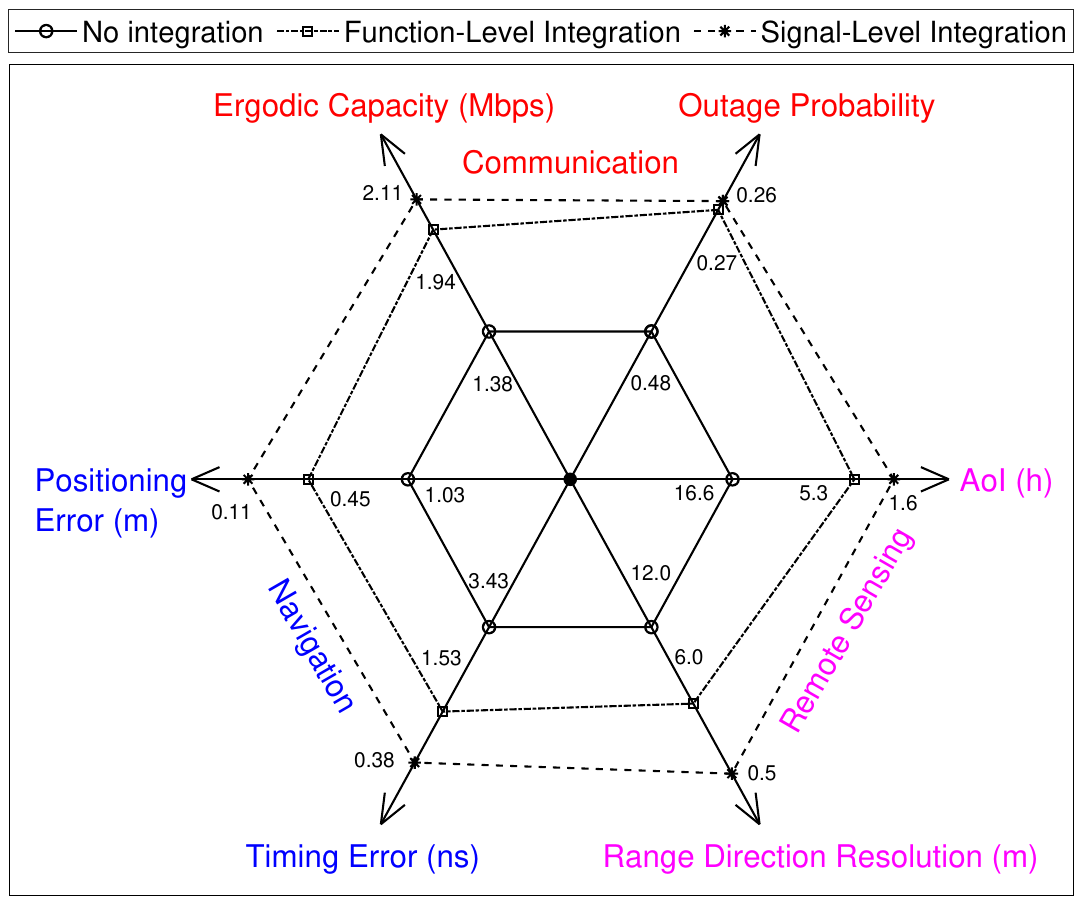}
\caption{Estimated gain of function-level and signal-level integration. \label{Fig:4}}
\end{figure}

The overall performance of the CNR is presented by a spider web shown in  Fig. 4, where the communication is evaluated by capacity outage probability  and mean ergodic capacity marked in red, the navigation is characterized by positioning accuracy  and timing error marked in blue, and the remote sensing is captured by resolution in range direction and AoI measured at the ground receivers marked in purple.
It can be observed that both function-level and signal-level integration can comprehensively improve the CNR performance.
The function-level integration introduces more available satellites for  the three functionalities, which shortens the communication distance, provides more anchor location references for navigation, and decreases the revisit time for remote sensing.
The gain of the signal-level integration mainly comes from  expansion of the available spectrum for each functionality, which results in less improvement in communication regime as the increased amount of spectrum is smaller than the other two functionalities.
More importantly, near centimeter-level positioning  achieved by the signal-level integration is quite fascinating for autonomous vehicles to perform lane navigation.
Meanwhile, the decimeter-level earth observation accuracy is helpful to distinguish different vehicles on the ground, which can be leveraged to conduct sophisticated road traffic analysis and  transportation system optimization.
Such quantitative improvement of the navigation and remote sensing performance by even the lowest two levels of integration, e.g., decreasing the positioning error and discernible distance in view by one order of magnitude, shines  light on the future ICNR industry,  while communication performance is also expected to be enhanced significantly by deeply exploiting the navigation and remote sensing data in higher levels of integration.
Finally, it is safe to admit the promising potentials of the ICNR in comprehensively enhancing the CNR performance as the integration degree goes up to data-level and application-level, by deep inter-functionality knowledge exploitation and application's QoS requirement domain exploration.

\section{Future Research Directions}

Optimization of the ICNR system has to jointly take the three functionalities into account.
However, due to distinct working mechanisms of the three functionalities from one  another, there are some trending challenges and open research directions in the ICNR framework, especially for vehicular applications.

\subsection{Constellation Design}

As the constellation size increases, the quality of communication may increase first due to shrinking coverage holes, and then decrease because of rising inter-satellite interference if no advanced interference coordination technique is implemented.
On the other hand, with the LEO constellation getting dense, the GDOP in the navigation regime greatly decreases to around 2 in the early stage, while less benefit is anticipated by deploying more satellites thereafter [11].
Furthermore, in addition to fast varying satellite distribution, the mobility of vehicles makes terminals experience more complicated positioning performance influenced by the constellation.
For the remote sensing functionality, the constellation size has a direct influence on the temporal resolution in terms of revisit time, where the revisit time in stripmap mode is generally  decreasing with the number of satellites.
It is worth noting that on-demand spotlight mode may be more suitable for transportation remote sensing for vehicular applications, where urban area has a higher requirement than rural area.
Since a mega-constellation of LEO satellites must be launched  batch by batch lasting for several years or more, it is thus critical to optimize the evolution route of the ever-growing constellation, so as to march in lockstep with the increasing CNR demand of autonomous driving.

\subsection{Satellite Association}

Associating a terminal  to the nearest access point has long been the fundamental principle in wireless communication field to experience the lowest channel fading, implying that the nadir of a satellite on the ground gains the best channel quality.
However, the thing becomes totally different for the navigation and remote sensing.
In particular, directly looking below with a view angle of $0^{\circ}$ fails to distinguish any two objects with any distance apart in SAR based remote sensing.
Therefore, a SAR satellite needs to take a side look, and a view angle of $25^{\circ}$ at an orbit altitude of $800$ km could lead to a distance of around 373 km between the swath center and the nadir.
Although the ground range resolution gets better as the view angle becomes larger,  a moderate view angle is usually adopted to avoid severe signal attenuation.
As for the navigation, the GDOP becomes better if the satellites are more scattered throughout the sky, indicating that the navigation functionality prefers to use signals emitted from far satellites dispersed in all directions.
In other words, the navigation terminals may need to discard some signals from nearby satellites  although many satellites can be seen in the era of mega-constellations of LEO satellites.
The distinct association preference among different functionalities, as shown in Fig. 5, causes great difficulty in multi-purpose waveform design and beam control towards moving vehicles in the proposed ICNR networks.
Additionally, on the terminal side, vehicles need to dynamically conduct satellite selection for each functionality to adapt to satellite distribution and own location variations.

\begin{figure}[!t]
\centering
\includegraphics[width=0.47\textwidth]{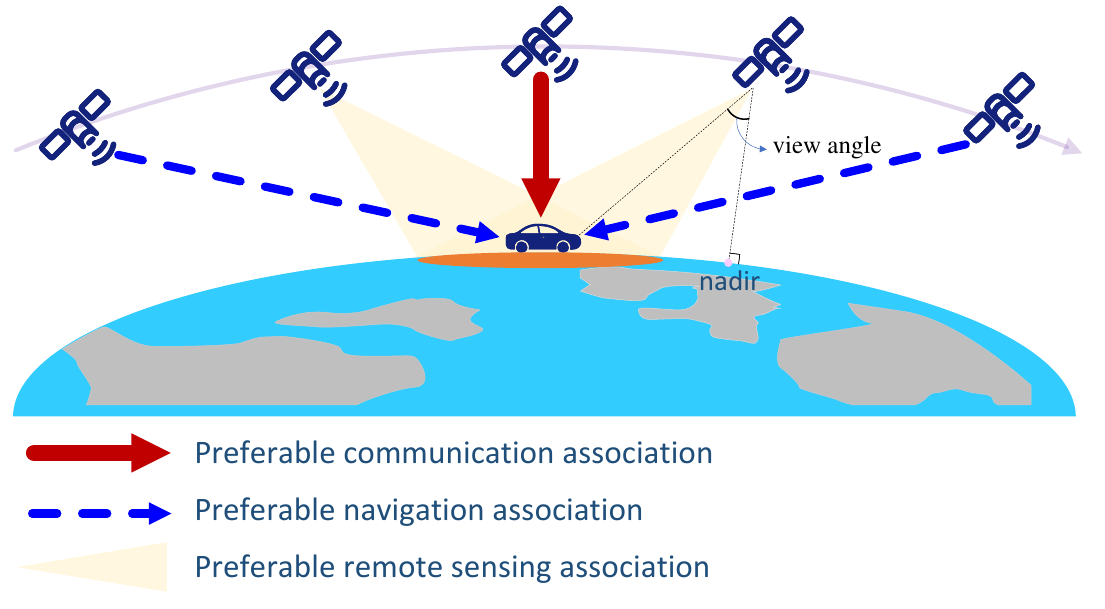}
\caption{ Preferable satellite association relationship for CNR functionalities. \label{Fig:5}}
\end{figure}

\subsection{Resource Management}

On-demand service providing could probably be one of the main features of the LEO mega-constellation based ICNR systems.
Particularly, as a small number of satellites are deployed in traditional remote sensing, the stripmap mode is commonly implemented to obtain earth observation images with global coverage, usually causing long revisit time greater than a couple of days or more.
By contrast, since every ground point is able to see many satellites in the LEO mega-constellations, the spotlight mode is more appropriate to be performed whenever there is a particular demand on real-time observation towards a region,  while the stripmap mode can be still conducted by a number of satellites to offer regular earth observation.
Therefore, how many satellites that work in the stripmap mode and how many that are reserved for sporadic services should be carefully designed.
In addition, as the CNR functionalities share the limited resources, such as power, computing, cache, and spectrum, smart allocation of these heterogeneous resources plays a prominent role in enhancing overall performance, including  how to balance communication and computing payload by appropriately processing the remote sensing data in orbit.
Since a growing number of fast moving vehicles persistently impose comprehensive CNR requirements,  sequential resource management decisions have to be optimized in response to user mobility and satellite topology variation, which is pretty challenging in the highly dynamic ICNR networks.

\section{Conclusion}

In this article, we have proposed the ICNR framework in LEO satellite networks with vehicular applications.
After presenting the most fundamental principles of the CNR functionalities, we have classified the proposed ICNR networks in different integration levels.
The function-level and signal-level integration have then been instantiated as a numerical case study to show the integration gain.
Finally, possible future research directions for more efficient exploitation of the ICNR framework have been highlighted.



\end{document}